\documentclass[final,3p,times,twocolumn]{elsarticle}

\usepackage{amsmath}
\usepackage{amsfonts}
\usepackage{amssymb}
\usepackage{graphicx} 
\usepackage{subcaption}
\usepackage{epstopdf}
\usepackage{bm}
\usepackage{lineno}

\usepackage{color, soul}
\usepackage[table,xcdraw]{xcolor}

\usepackage{multirow}
\usepackage{booktabs}
\biboptions{numbers,sort&compress}
\usepackage{hyperref}
\usepackage{gensymb}
\usepackage[labelformat=empty, position=top]{subcaption}
\usepackage[export]{adjustbox}
\usepackage{textcomp, gensymb}
\usepackage{float}
\hypersetup{colorlinks,breaklinks,
            urlcolor=Maroon,
            linkcolor=Maroon}
\hypersetup{pdfauthor={Name}}
\journal{Science Advances}
\usepackage{amsmath,array, amssymb}
\usepackage{siunitx,mhchem}
\usepackage{blindtext}
\usepackage{overpic}

\begin{document}
\begin{frontmatter}

\title{Microstructure-Aware Bayesian Materials Design} 

\author{Danial Khatamsaz$^{a,*}$, Vahid Attari$^{a}$, Raymundo Arr\'{o}yave$^{a,b,c}$}
\address{$^{a}$Materials Science and Engineering Department, Texas A\&M University, College Station, TX, USA 77843}
\address{$^{b}$Mechanical Engineering Department, Texas A\&M University, College Station, TX, USA 77843}
\address{$^{c}$Department of Industrial \& Systems Engineering, Texas A\&M University, College Station, TX 77843}


\begin{abstract}
In this study, we propose a novel microstructure-sensitive Bayesian optimization (BO) framework designed to enhance the efficiency of materials discovery by explicitly incorporating microstructural information. Traditional materials design approaches often focus exclusively on direct chemistry-process-property relationships, overlooking the critical role of microstructures. To address this limitation, our framework integrates microstructural descriptors as latent variables, enabling the construction of a comprehensive process-structure-property mapping that improves both predictive accuracy and optimization outcomes. By employing the active subspace method for dimensionality reduction, we identify the most influential microstructural features, thereby reducing computational complexity while maintaining high accuracy in the design process. This approach also enhances the probabilistic modeling capabilities of Gaussian processes, accelerating convergence to optimal material configurations with fewer iterations and experimental observations. We demonstrate the efficacy of our framework through synthetic and real-world case studies, including the design of Mg$_2$Sn$_{x}$Si$_{1-x}$ thermoelectric materials for energy conversion. Our results underscore the critical role of microstructures in linking processing conditions to material properties, highlighting the potential of a microstructure-aware design paradigm to revolutionize materials discovery. Furthermore, this work suggests that since incorporating microstructure awareness improves the efficiency of Bayesian materials discovery, microstructure characterization stages should be integral to automated—and eventually autonomous—platforms for materials development.
\end{abstract}


\end{frontmatter}

\section{Introduction}
\label{sec:intro}

\begin{figure*}[!ht]
    \centering
    \includegraphics[width=0.98\textwidth]{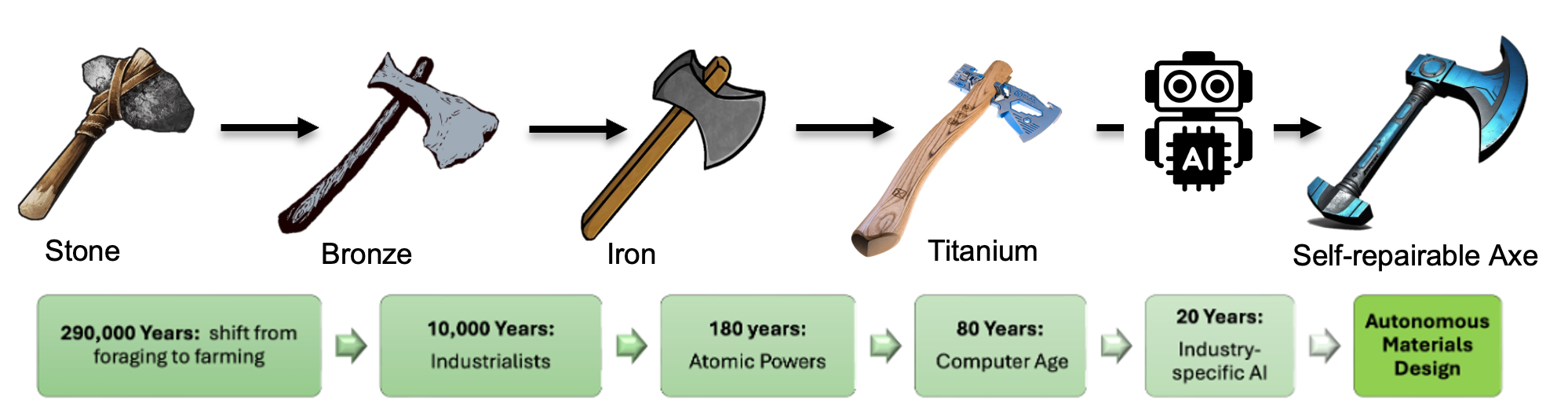}
    \caption{\textbf{Throughout history, material innovations have been a driving force behind societal and technological advancements, from the Stone Age to the AI era.} As humanity faces modern challenges like climate change and resource scarcity, the need for novel materials and accelerated development approaches is more urgent than ever. AI-driven autonomous labs and robotics will accelerate materials development, creating a ‘Moore’s law for research’ that will potentially fast-track breakthroughs in clean energy and beyond.}
    \label{GA}
\end{figure*}

Materials development has been central to societal progress, marking key historical transitions such as the Stone Age, Bronze Age, Iron Age, and the Industrial Revolution. The names of these eras highlight the transformative impact of new materials on human history. It took nearly hundreds of thousands of years to transition from hunting and gathering to farming~\cite{bocquet2011world}, about 10,000 years to industrialize, and only 180 years to harness atomic power~\cite{mokyr1992lever,rhodes2012making}. The computer age began within the last century, with rapid advancements over the past 50 years, and the development of artificial intelligence has arguably \emph{accelerated the rate of acceleration} in innovation further in the past two decades~\cite{vaswani2017attention, devlin2018bert}. Each of these inflections in technology development---see Fig.~\ref{GA}---have come accompanied by an enhanced ability to control the structure of matter at increasingly finer scales, transforming society along the way~\cite{brynjolfsson2014second}.

Today, the urgent need for novel materials to address critical challenges such as climate change, resource scarcity, and sustainable energy~\cite{hoffert2002advanced} requires a fundamental shift in how materials are discovered and optimized. The traditional trial-and-error approach to material development, while effective in the past, is increasingly inadequate for meeting the accelerated pace of innovation demanded by these global challenges~\cite{jain2016computational,butler2018machine}. A key limitation of current methods lies in their treatment of microstructural information. Although advanced techniques now allow microstructures to be characterized across many length scales, the materials development process remains largely microstructure-agnostic. Microstructures are rarely treated as direct design targets; instead, they are viewed as emergent by-products of composition and processing choices. For example, in the design of advanced alloys, engineers typically optimize elemental compositions and heat-treatment parameters to achieve specific properties, such as strength or toughness. However, the resulting microstructural features, such as grain boundaries or phase distributions, are only analyzed \emph{after} fabrication, and no information about materials microstructure enters the decision-making process during materials development tasks. 

Microstructures form the critical link between chemistry, processing protocols, and the resulting properties and performance of materials~\cite{himanen2019data,fullwood2010microstructure,white2023digital,reed2008superalloys,raabe2019strategies}. This relationship is encapsulated in the Process-Structure-Property-Performance (PSPP) chain, a foundational framework in materials science that describes how processing methods lead to specific microstructures, which in turn determine material properties and performance\cite{khosravani2017development,kalidindi2015materials}. For example, in thermoelectric materials, microstructure strongly influences the efficiency of converting heat into electricity. Fine-tuning grain size, phase distribution, and defect concentration can enhance performance by reducing thermal conductivity while improving electrical conductivity~\cite{snyder2008complex,zebarjadi2012perspectives,xue2016accelerated,lookman2019active}. 

Despite its central role in the PSPP paradigm, microstructure is traditionally treated as an emergent by-product of composition and processing rather than as a directly tunable design parameter. This limitation is partly due to current constraints in bottom-up materials assembly, where atomic-level control is not yet feasible for most materials systems---an exception is meta-materials, which can be designed and fabricated with specific, \emph{addressable}, macroscopic unit cells\cite{ha2023rapid}. However, explicitly incorporating microstructure as a controllable (or at least addressable) element in the decision-making process~\cite{arroyave2019systems} during materials development could unlock new pathways for high-performance materials design.

Recently, the community has begun to introduce microstructure-aware approaches in materials optimization. For instance, Morand \emph{et al.}~\cite{morand2024machine} and Flores Ituarte \emph{et al.}~\cite{flores2023optimisation} used multi-objective and machine-learning methods to correlate process variables, microstructural features, and final properties. In a physics-plus-data setting, Liao \emph{et al.}~\cite{liao2024process} modeled freeze-cast ceramics and highlighted the computational demands of bridging the PSP chain. On the stochastic side, Tran and Wildey~\cite{tran2021solving} inferred distributions of microstructural design variables to achieve target property distributions, whereas Wu and Hufnagel~\cite{wu2024efficient} inverted heat-treatment schedules for aluminum alloys using Bayesian Optimization (BO) and evolutionary strategies. Under limited-data constraints, Rixner and Koutsourelakis~\cite{rixner2022self} employed a self-supervised mechanism to refine surrogate models more selectively. Despite these advances, many methods still view microstructures as \emph{post-facto} outcomes or assume deterministic PSPP relationships that do not account for uncertainties in the PSPP chain. Moreover, truly \emph{practical} microstructure-driven design typically should be able to operate in a sparse-data regime, making Bayesian formulations especially appealing. By quantifying predictive uncertainty and prioritizing high-value data—even when data are scarce—Bayesian approaches could provide a robust path to microstructure-centric design. In this context, a subset of the present authors~\cite{molkeri2022importance}  showed that microstructure features could accelerate the performance of otherwise conventional Bayesian materials discovery workflows, although in that work the microstructure feature responsible for the property of interest was known \emph{a priori} and was not discovered during the campaign, limiting the usefulness of the approach. 

While the examples of microstructure-aware materials development have been tested and deployed \emph{in silico}, their impact will likely be greater in campaigns focused on optimizing 'real' materials. Materials Acceleration aims to drastically reduce the materials development cycle to 1–2 years by leveraging AI-driven methods transforming R\&D into a rapid, data-driven process~\cite{aspuru2018materials,agrawal2016perspective,raccuglia2016machine,xue2016accelerated,lookman2019active,pilania2013accelerating}. Often referred to as Materials Acceleration Platforms (MAPs) or Self-Driving Materials Laboratories~\cite{miracle2024autonomous}, these integrated systems combine high-throughput experiments, computational modeling, and data analytics to accelerate materials discovery at scale~\cite{aspuru2018materials,miracle2024autonomous}, and potentially establishing a “Moore's law for research,” accelerating the rate of innovation in materials~\cite{liu2017materials}. Central to this paradigm is not just the speed of discovery, but also the quality and performance of the materials being developed. 

Although these methods have shown promise at optimizing compositional and processing parameters to attain optimal property/performance metrics, the potential of improving the efficiency and effectiveness of the workflows by incorporating microstructure information in the design/discovery process remains largely unexplored. Our earlier work demonstrated\cite{molkeri2022importance} the importance of microstructural information in materials design but required knowing \emph{a priori} which features were critical, leaving open the need for a more general, systematic integration of microstructure into BO-driven frameworks. This paper explores whether extending BO to include microstructural descriptors (e.g., grain size, phase fractions, defect metrics) is both technically feasible and beneficial to the outcomes of materials discovery. 

The incorporation of microstructure information in Bayesian discovery workflows requires significant modifications to traditional BO schemes. In a Bayesian-based design framework, the true computational or experimental model is represented by a surrogate model, often a Gaussian process (GP), which constructs mappings from the design space to the objective space~\cite{Rasmussen:2005:GPM:1162254,costabal2019multi}. The performance of BO relies on the GP’s probabilistic modeling of objective functions, treating systems as black boxes that only consider input/output data without needing intermediate variables. Decision-making is driven by a utility function evaluating expected gains from potential experiments. This framework is sometimes described as “latent space agnostic,” meaning it does not explicitly model intermediate variables, even though relationships can exist between these hidden variables and the input/output data~\cite{hastie2009elements}. Intermediate variables, such as microstructure-related information, are not directly observed but are inferred from measured variables and represent underlying processes affecting the observed data. Extracting and exploiting these relationships can potentially enhance the design framework by reducing the complexity of learning how design variables map to objective functions~\cite{Rasmussen:2005:GPM:1162254, jones1998efficient}.

In this work, we demonstrate a \emph{Latent-Space-Aware BO} scheme that integrates microstructural information into the materials design framework. Our approach exploits PSPP relationships in a Bayesian setting, a concept not extensively explored in previous works except for Molkeri et al.~\cite{molkeri2022importance}. This new BO framework leverages microstructural features as latent variables to enhance input–output mappings and improve the design process. In materials design, this translates to a microstructure-sensitive BO, where these features connect chemistry and process parameters to material properties, thereby improving the GP’s probabilistic modeling capabilities and overall design performance. Unlike traditional methods, our approach can handle a larger number of features, surpassing limitations in prior research~\cite{molkeri2022importance}. To effectively utilize these latent variables, it is crucial to determine their impact on the objective function’s variability. We implement the \emph{Active Subspace Method} as a dimensionality reduction technique to identify key subspaces within the latent space that influence objective function variability~\cite{constantine2014active,constantine2014computing,constantine2015active}. These subspace components create a reduced yet informative representation of the latent space, in turn enhancing GP modeling accuracy and design performance in higher-dimensional spaces. 

To highlight the importance of incorporating these latent variables, we first demonstrate our method on a synthetic mathematical example. We then apply the proposed framework to a real-world materials design problem involving microstructural descriptors, underscoring the critical role of microstructures in linking processing conditions to material properties. Our findings suggest a paradigm shift from the traditional chemistry/process–property approach to a more integrated chemistry/process–structure–property methodology.

While our focus here is the application of latent-space-aware BO for accelerating materials design using microstructure-sensitive methods, this methodology can potentially be integrated with robotic systems and Self-Driving Laboratories (SDLs) to further reduce development times~\cite{abolhasani2023rise,seifrid2022autonomous,miracle2024autonomous}. By coupling real-time microstructure analysis with autonomous experimental capabilities and the computational framework proposed here, we envision a transformative acceleration in materials discovery. This integration not only automates complex tasks but also refines precision and speed in finding new materials, potentially revolutionizing the cycle from concept to application.

\begin{figure*}[!ht]
    \centering
    \includegraphics[width=1\textwidth]{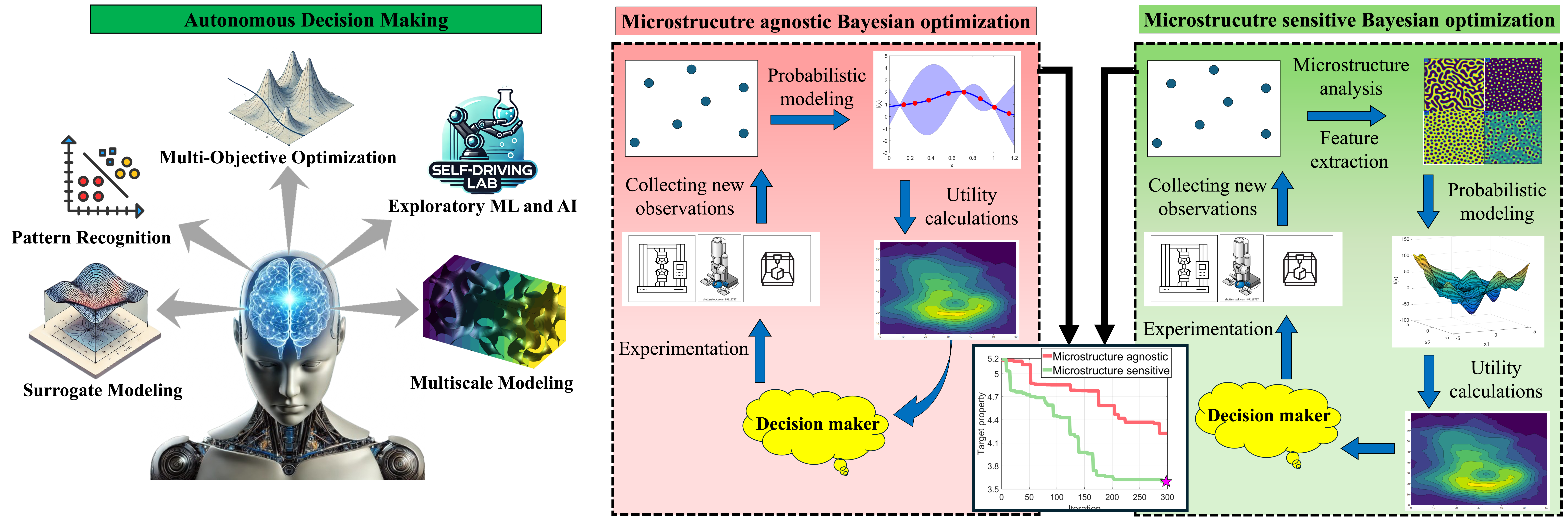}
    \caption{(Left) Autonomous materials decision-making using microstructure-sensitive design methods, and (Right) the main steps of the microstructure-sensitive Bayesian optimization framework. The approach extracts property-relevant information from microstructural features via dimensionality reduction and conditions a Gaussian process on both design variables and reduced microstructure representations, thereby learning the chemistry–structure–property relationship.}
    \label{GA2}
\end{figure*}

In Section~\ref{sec:methods}, we detail our microstructure-aware BO framework, including how latent variables and dimensionality reduction are leveraged. Section~\ref{sec:results} presents both a synthetic test problem and a real-world case study, demonstrating the performance gains achieved by explicitly modeling microstructures. We conclude in Section~\ref{sec:conclusions} with a discussion of potential integrations into SDLs and the broader implications for rapid materials development.

\section{Methods}
\label{sec:methods}

Here, we describe a computational framework capable of integrating microstructural descriptors into iterative Bayesian materials design schemes. The framework combines chemistry and processing parameters with microstructure-derived features to create an augmented design space, enabling predictions that capture subtle process-structure-property-performance (PSPP) relationships. A Gaussian Process (GP) regression model is used to predict material properties in unexplored regions of the design space while accounting for uncertainties. To handle the increased dimensionality, the framework employs an active subspace method to reduce the design space into a lower-dimensional, highly informative representation that retains critical variability for optimization. Bayesian Optimization (BO) is then applied to iteratively refine the design, dynamically incorporating new data and updating the active subspace.

After testing the framework against a synthetic mathematical problem, we apply it to tackle a computational materials design problem consisting of minimizing the thermal conductivity of a virtual two-phase thermoelectric microstructure. Simulations are performed using a CALPHAD-reinforced elasto-chemical phase-field method to model microstructure evolution, and micro-elasticity modeling is employed to capture stress-strain interactions at the microscale. Microstructural descriptors, such as phase fraction, Fourier scattering intensity, and Shannon entropy, quantify the structural complexity and are incorporated into the optimization process as features used to augment the chemistry-processing input (design) space. Heat transfer simulations assuming Fourier’s law are used to calculate the thermal conductivity of the microstructure, which is then used as the target for the optimization.

\subsection*{Introduction to the Framework}
As illustrated in Figure~\ref{GA}, our proposed framework integrates microstructure information into the design process by generating microstructure descriptors from the chemistry and processing parameters. These descriptors are then used to expand the input space, thereby forming an augmented design space as initially proposed in earlier work~\cite{molkeri2022importance}. A GP is conditioned on this merged space of design variables and projected (intermediate) descriptors to predict properties in unexplored regions. In order to address the increased dimensionality of the input space, the augmented design space is then projected into an active subspace~\cite{khatamsaz2021adaptive}, which is a lower-dimensional but highly informative space that captures most of the variability in the objective function. An acquisition function then guides the selection of the next experiment based on the information derived from this reduced space, and the next design space is determined by projecting back to the original chemistry-processing (controllable) input space. Below, we describe some of the key components of the \emph{Latent-Space-Aware BO} scheme.

\subsection{Incorporating latent space information into BO}

In goal-oriented materials design, the focus is typically on identifying the optimal chemistry and processing conditions to achieve target properties/performance metrics. Traditionally, the design process is agnostic to microstructure details, favoring more direct/pragmatic approaches that exploit (chemistry-)process-property (PP) relationships directly~\cite{molkeri2022importance}. However, latent variables, such as microstructure feature descriptors, can be crucial for capturing subtle relationships between process, structure, and property. Therefore, we introduce the active subspace method, which effectively captures these relationships by forming principal components that represent the largest variability in the objective function within the feature space. These principal components are derived from the eigenvectors of a covariance matrix calculated using the local gradients of the objective function in the latent space at observed locations---here we note that such dimensionality reduction is different from other approaches, such as conventional Principal Component Analysis (PCA)\cite{abdi2010principal} as in this case the reduction takes into account both the input and output spaces. The associated eigenvalues indicate the extent to which the objective function varies along each eigenvector. By normalizing these eigenvalues to sum up to 1, we establish a criterion for determining the fraction of variation information that should be captured from the latent space. Setting this threshold to 1 means selecting all eigenvectors to form the active subspace, which maintains the same dimensionality as the original latent space but with a transformed coordinate system. Conversely, a lower threshold results in selecting only the essential eigenvectors that capture a significant portion of the variability.

In our latent-space-aware BO framework, the GP is conditioned on the combined space of design variables and the active subspace of latent variables. The standard BO process is then followed by iteratively proposing experiments and updating the model. Notably, the active subspace is dynamically updated as new observations are integrated, ensuring that the framework adapts to the evolving data landscape.

\subsection{Gaussian process regression}

Gaussian processes (GPs) are commonly used probabilistic models to represent underlying objective functions in BO frameworks due to the capability to provide prediction uncertainty, flexibility in modification and manipulation, and low computational cost~\cite{Rasmussen:2005:GPM:1162254}. 
A GP is conditioned on $N$ previously observed data denoted by $\{\mathbf{X}_{N}, \mathbf{y}_{N}\}$, where $\mathbf{X}_{N} = (\mathbf{x}_{1}, \ldots, \mathbf{x}_{N})$ and $\mathbf{y}_{N} = \left(f(\mathbf{x}_{1}), \ldots, f(\mathbf{x}_{N})\right)$, and it represents the prediction at an unobserved location \(\mathbf{x}\) by a normal distribution:
\begin{equation}
\label{GP11}
f_{\textrm{GP}}(\mathbf{x}) \mid \mathbf{X}_{N}, \mathbf{y}_{N} \sim \mathcal{N}\left(\mu(\mathbf{x}),\sigma_{\textrm{GP}}^2(\mathbf{x})\right)
\end{equation}
where
\begin{equation}
\begin{aligned}
\label{meancov}
\mu(\mathbf{x}) &= K(\mathbf{X}_{N},\mathbf{x})^\textrm{T}[K(\mathbf{X}_{N},\mathbf{X}_{N})+\sigma^2_{n}I]^{-1} \mathbf{y}_{N}\\
\sigma_{\textrm{GP}}^2 (\mathbf{x})  &=  k(\mathbf{x},\mathbf{x}) - K(\mathbf{X}_{N},\mathbf{x})^\textrm{T} [K(\mathbf{X}_{N},\mathbf{X}_{N})+\sigma^2_{n}I]^{-1}K(\mathbf{X}_{N},\mathbf{x})
\end{aligned}
\end{equation}
with $k$ as a real-valued kernel function, $K(\mathbf{X}_{N},\mathbf{X}_{N})$ as an $N \times N$ matrix with the $(m,n)$ entry as $k(\mathbf{x}_{m}, \mathbf{x}_{n})$, and $K(\mathbf{X}_{N}, \mathbf{x})$ is an $N \times 1$ vector with the $m^{th}$ entry as $k(\mathbf{x}_{m}, \mathbf{x})$.  $\sigma^2_{n}$ models observation error, if any. A commonly used kernel function is the squared exponential:
\begin{equation}
\label{eq:5}
k(\mathbf{x},\mathbf{x'}) = \sigma_s^2 \exp\left(- \sum_{h = 1}^{d} \frac {(x_h-x'_h)^2}{2l_h^2}\right )
\end{equation}
where $d$ is the input space dimensionality, $\sigma_s^2$ is the signal variance, and $l_h$, where $h = 1,2,\ldots,d$, is the characteristic length-scale that determines how much correlation exists between observations within dimension $h$.

\subsection{Active subspace method}

The active subspace method (ASM) is a dimensionality reduction technique to construct a lower-dimensional representation of an objective function while preserving as much information as possible about the objective function variability~\cite{ghoreishi2019adaptive,russi2010uncertainty,khatamsaz2021adaptive}. This method is key to our framework, as it streamlines the input space without sacrificing the key variations that drive performance. The constructed lower-dimensional space is referred to as the active subspace, with its principal components being linear combinations of the original space that induce the largest objective function variability.

Following Refs. \cite{constantine2014active,ghoreishi2019adaptive,constantine2014computing}, assuming a scalar function \(f(\mathbf{x})\) is defined on the input space \(\mathcal{X} \subset \mathbb{R}^m\), the gradient \(\nabla_{\mathbf{x}} f\) represents its sensitivity at \(\mathbf{x}\). The covariance of the gradient, \(\mathbf{C}\), is defined as
\begin{equation}
\mathbf{C} = \mathbb{E} [\nabla_\mathbf{x} f(\mathbf{x}) \nabla_\mathbf{x} f(\mathbf{x})^\textrm{T}]
\end{equation}

\begin{figure}[]
    \includegraphics[width=0.45\textwidth]{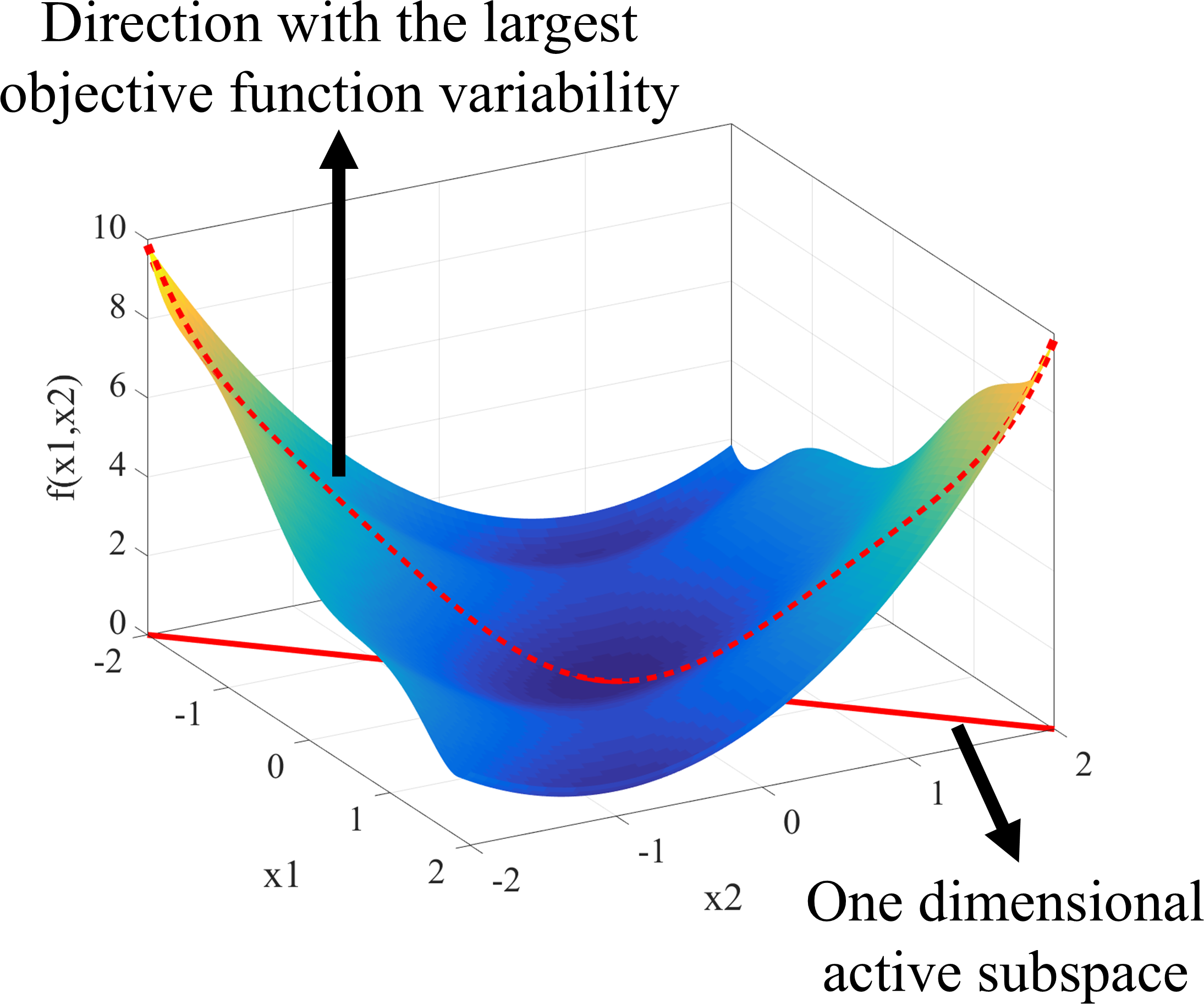}
    \caption{\textbf{Illustration of the active subspace definition over a 2-dimensional input space.} The active subspace determines the subspace that represents the largest objective function variability (the most active direction) to build a lower-dimensional projection of the function. In the example above, the active subspace is represented as 1-dimensional subspace that can be traversed by a scalar.}
    \label{active_sub}
\end{figure}

Assuming that there are \(M\) previously evaluated samples from the function, the covariance matrix is approximated as
\begin{equation}
\label{monte}
    \mathbf{C} \approx \frac{1}{M} \sum_{i=1}^M \nabla_\mathbf{x} f(\mathbf{x}_i) \nabla_\mathbf{x} f(\mathbf{x}_i)^\textrm{T}
\end{equation}
The eigenvectors of the covariance matrix are taken as new principal components of the space, and their corresponding eigenvalues indicate their importance in objective function variability. Based on the eigenvalue decomposition, the covariance matrix can be written as
\begin{equation}
\label{decomp}
    \mathbf{C} = \mathbf{W}\zeta \mathbf{W}^\textrm{T}
\end{equation}
where \(\mathbf{W}\) is the matrix of eigenvectors and \(\zeta\) is a diagonal matrix of the respective eigenvalues. By picking the eigenvectors associated with the \(n\) largest eigenvalues, an \(n\)-dimensional active subspace is defined. 
\begin{equation}
\label{divider}
    \mathbf{W} = [\mathbf{U} \; \; \mathbf{V}], \quad
\zeta = \begin{bmatrix}
\zeta_1 & \\
& \zeta_2 \\
\end{bmatrix}
\end{equation}
The matrix \(\mathbf{U}\) contains the \(n\) eigenvectors corresponding to the largest eigenvalues (forming \(\zeta_1\)) and is called the transformation matrix. Any point in \(\mathcal{X}\) can be projected to the active subspace using
\begin{equation}
\label{transformation}
    \mathbf{z}=\mathbf{U}^\textrm{T}\mathbf{x}
\end{equation}
and the function \(g\) represents the original function \(f\) in this active subspace as
\begin{equation}
    g(\mathbf{z}) = g(\mathbf{U}^\textrm{T}\mathbf{x}) \approx f(\mathbf{x})
\end{equation}
Once the transformation matrix \(\mathbf{U}\) is determined, all evaluated points from \(f\) are projected to the active subspace:
\begin{equation}
    \mathbf{Z}_N = \mathbf{U}^\textrm{T}\mathbf{X}_N
\end{equation}
to obtain a lower-dimensional representation of the data. A 2-dimensional example is shown in figure~\ref{active_sub} where the active subspace is essentially a 1-dimensional line.

\subsection{Microstructure Modeling: CALPHAD Reinforced Elasto-chemical Phase-field Method}\label{sec:}

Phase-field modeling, a computational method for simulating microstructure evolution in materials, represents interfaces as diffuse regions rather than sharp boundaries~\cite{chen2022classical}. Transitioning now to the simulation of microstructures, this method serves as the basis for modeling the evolution of the microstructure under various processing conditions. It is a versatile tool for investigating diverse phenomena such as grain growth, solid-state phase transformations, ferroelectric phase transitions, electrochemical processes, and mechanical deformation including dislocation motion, twinning, and fracture. The local mass flux in the presence of a composition gradient, represented by the diffusion flux \(\mathbf{J}\) (in units of \(\si{mol~m^{-2}~s^{-1}}\)), is described by the linear kinetic theory equation:
\begin{equation}
\mathbf{J} = -\mathbf{M} \nabla \mu^{tot}
\end{equation}
where \(\mathbf{M}\) is the interface mobility (assumed isotropic) and \(\mu^{tot}=\frac{\delta \mathcal{F}^{tot}}{\delta c}\) is the total potential driving the kinetic transition. The Cahn-Hilliard (C-H) kinetic equation, integral to studying microstructure evolution, is expressed as:
\begin{equation} \label{eqn:C_Heqn}
\frac{\partial c}{\partial t} = \nabla\cdot \mathbf{M} \nabla \left ( \frac{\delta \mathcal{F}^{tot}}{\delta c} \right )
\end{equation}
This equation is solved using a semi-implicit Fourier spectral approach in frequency space~\cite{chen1998applications}. The simulation employs a \(512\times512\) cell size, with composition profiles perturbed randomly by \(\pm2\%\) around the alloy composition using a constant seed number for consistency. A constant time step and regular data-saving intervals are maintained throughout to ensure comparability and stability of simulations under varying kinetic parameters, which is crucial for understanding coarsening rates.

\subsection{Micro-elasticity model}\label{sec:microelasticity}

The micro-elasticity problem pertains to the study of long-range interactions in materials at the microscale, focusing on the relationship between stress, strain, and material properties. Building on the simulated microstructure, we next model the mechanical interactions using a micro-elasticity approach. It aims to characterize the elastic response of materials at the smallest scales, considering the heterogeneous nature of the material and its constituents. The micro-elasticity equations are given as~\cite{attari2019exploration}:
     
\begin{equation} \label{eqn:mech_eq}
    \frac{\partial \sigma_{ij}}{\partial r_j} = 0 \quad \text{in} \quad \Omega
\end{equation}     
\begin{equation} \label{eqn:el_strain}
    \varepsilon_{ij} = \frac{1}{2} \left ( \frac{\partial u_i}{\partial r_j} - \frac{\partial u_j}{\partial r_i} \right )  
\end{equation}     
\begin{equation} \label{eqn:hookslaw}
    \sigma_{ij} = C_{ijkl}\varepsilon^{el}_{kl} 
\end{equation}
\begin{equation} \label{eqn:eps_tot}
    \varepsilon^{el}_{kl} = \varepsilon^{tot}_{kl} - \varepsilon^{0}_{kl} 
\end{equation}    
\noindent
where Equations~\ref{eqn:mech_eq}--\ref{eqn:eps_tot} represent mechanical equilibrium, kinematics, Hooke's law, and the strain relationship, respectively. \(u\) denotes the displacement field and \(r\) represents the spatial coordinate. The dilatational eigenstrain term, \(\varepsilon^{0}_{kl}\), is expressed as \(\varepsilon^{T}\delta_{kl}h(c)\), accounting for lattice strain between phases. Here, \(\varepsilon^{T}\) signifies the strength of the eigenstrain, \(\delta_{kl}\) denotes the Kronecker delta, and \(h(c)=c^3(10-15c+6c^2)\) is a standard scalar interpolation function. The composition-dependent fourth-order elastic modulus tensor, \(C_{ijkl}\), is described as:
\begin{equation}
C_{ijkl}(c) = C^{eff}_{ijkl} - g(c)\Delta C_{ijkl}
\end{equation}
Here, \(\Delta C_{ijkl} = C^{\alpha}_{ijkl} - C^{\beta}_{ijkl}\) represents the difference between the elastic moduli tensors of phases \(\alpha\) and \(\beta\). In the case of linear elasticity, \(C_{ijkl}\) is simplified to two independent constants due to cubic phase symmetry. The microelasticity problem is solved using the FFT-based iterative method, accounting for stress-free transformation strains for each phase, inhomogeneous elastic constants, strain-control based on stress-control, and periodic boundary conditions. Convergence is reached when the \(L^2\) norm, \(||u^{n+1}-u^{n}||\), is less than \(10^{-8}\).

\subsection{Microstructure Representation}

Having simulated the microstructure and its mechanical behavior, we now focus on quantitatively representing the microstructure using various physical descriptors. In this study, we utilize physical microstructure descriptors, including phase fraction, phase composition, radially-averaged Fourier scattering intensity of microstructures, and Shannon entropy. Phase fraction estimation involves computing a global image threshold using Otsu's method~\cite{otsu1979threshold} from grayscale images. Phase composition is derived by identifying maximum and minimum composition values within the domain. The radially averaged Fast Fourier Transform (FFT) of the field order parameters serves as a measure of the characteristic length scale in the microstructure domain. Calculations reveal distinctive curves for each image at a fixed time, with the curve and its peak shifting towards the right-hand side as the annealing time increases (Refer to~\cite{honarmandi2022accelerated}). Shannon entropy, when applied to a microstructure, measures the randomness present in the distribution of composition intensities within the microstructure. It quantifies the amount of information required to describe the microstructure content. This measure provides insight into the complexity or disorder of the microstructure, with higher entropy values indicating greater variability in composition. The concept delineates interfaces in the microstructure where the greatest variability occurs in the transition from one phase to another. Mathematically, Shannon entropy is calculated using the probability distribution of composition intensities within the microstructure.

\subsection{Effective Thermal Conductivity}\label{sec:method_heat}

Finally, to complete the feedback loop in our design framework, we link the microstructural characteristics to a performance metric by analyzing the effective thermal conductivity. The homogenized equations, with appropriate boundary or initial conditions, are used to analyze steady-state heat transfer across various microstructures. Effective properties of random heterogeneous materials are determined through ensemble averaging of local fields governed by relevant conservation equations, such as partial differential equations. The Fourier heat conductivity equation with a heterogeneous coefficient is solved in the steady-state case to simulate heat flow through computed microstructures. The heterogeneous thermal diffusivity is given by 
\[
\alpha=\phi\alpha_{p_{1}}+(1-\phi)\alpha_{p_{2}} + A_{GB}\phi^2(1-\phi^2)
\]
where \(\phi\) is the scaled order parameter (between 0 and 1) and \(A_{GB}\) controls the grain boundary thermal diffusivity. Constant heat-flux boundary conditions are applied on the microstructure in the direction of orthogonal heat flow from the bottom-left to the top-right corner. Thermal flow is driven by constant temperature difference boundary conditions across the structure in the primary flow direction, with a constant temperature used as the initial condition. For classical, isotropic materials, heat conduction is defined by Fourier’s law~\cite{toberer2012advances}:
\begin{equation}\label{eq:TC}
q = -\kappa \nabla T
\end{equation}
where \(\kappa\) represents the material's scalar thermal conductivity in SI units (Wm\(^{-1}\)K\(^{-1}\)) and \(\nabla T\) is the temperature gradient vector. The computation time needed to calculate effective thermal conductivity, meeting the defined tolerance, is approximately 250 seconds using 28 CPUs.

\section{Results and Discussion}
\label{sec:results}

In this section, we demonstrate the performance of our microstructure-aware Bayesian optimization framework through both synthetic and practical examples. First, we illustrate the method on a notional problem defined over a four-dimensional design space that is mapped to a higher-dimensional latent space. By comparing the latent space-agnostic BO with our latent space-aware BO under different information thresholds, we show that incorporating latent variables leads to improved convergence toward the optimal solution. Activity scores and the dimensionality of the active subspace are used to assess the global sensitivity of the objective function to individual latent variables, highlighting their influence on design performance. We then extend our investigation to a realistic chemistry–microstructure–property modeling problem, where our approach is used to optimize the effective thermal conductivity of multiphase materials. Together, these examples illustrate that explicitly incorporating microstructural (latent) information within a Bayesian optimization framework enhances both efficiency and accuracy.

\subsection*{Notional Demonstration with Synthetic Problems}

A maximization problem is defined as
\begin{equation}
\label{optimization}
    \text{Find} \quad \mathbf{x}^{*}= \arg\max_{\mathbf{x} \in \mathcal{X}} g(\mathbf{x})
\end{equation}
where \(g\) is the objective function and \(\mathbf{x}^{*}\) is the optimal solution. The created synthetic function is defined over a 4-dimensional design space, \(\mathbf{x}=[x_1,\, x_2,\, x_3,\, x_4]\), mapped to a 6-dimensional latent space, \(\mathbf{F}=[f_1,\, f_2,\, f_3,\, f_4,\, f_5,\, f_6]\), and finally a scalar function of the latent variables \(g(\mathbf{F})\). Note that the objective function can be directly written as a function of the design variables and the latent space-agnostic BO supposedly exploits that direct relationship to learn and discover the optimal solution. However, it is expected that it needs more observations to learn such a complex relationship between the design variables and the objective function. The latent space-aware BO, on the other hand, exploits information from the latent space in addition to the relationship between design variables and the objective function to construct a GP for probabilistic modeling of the objective function. To mimic the situation in a materials design problem with access to microstructural feature descriptors, we assume the 6th latent variable, \(f_6\), is not known and thus, not considered in the active subspace calculations. The reason is two-fold: first, to show that the framework can exploit information from only known microstructural features as there may be some features that are not yet discovered by scientists or are expensive to measure and second, we still maintain some dependency on the design variables. Note that we still use Eqs.~\ref{toy_features} and \ref{toy_target} to calculate the objective function but the objective function information is extracted from the remaining 5 latent variables.

\begin{equation}
\begin{aligned}
    f_1 &= x_1^2 + \exp\Big({\frac{-x_2}{x_3}}\Big)\\[1mm]
    f_2 &= x_1 + x_3\\[1mm]
    f_3 &= \frac{x_2}{1+x_3}\\[1mm]
    f_4 &= \log (x_4+1) \times x_1\\[1mm]
    f_5 &= x_2 \times \sin{x_4}+\exp({x_1})\\[1mm]
    f_6 &= \sin{x_3}+\cos{x_4}
\end{aligned}
\label{toy_features}
\end{equation}

\begin{equation}
    y = f_1\times f_2 + \frac{f_2}{f_3} + f_5 \times f_4 + f_6
    \label{toy_target}
\end{equation}

\begin{figure*}[!ht]
    \centering
    \begin{overpic}[width=0.45\textwidth]{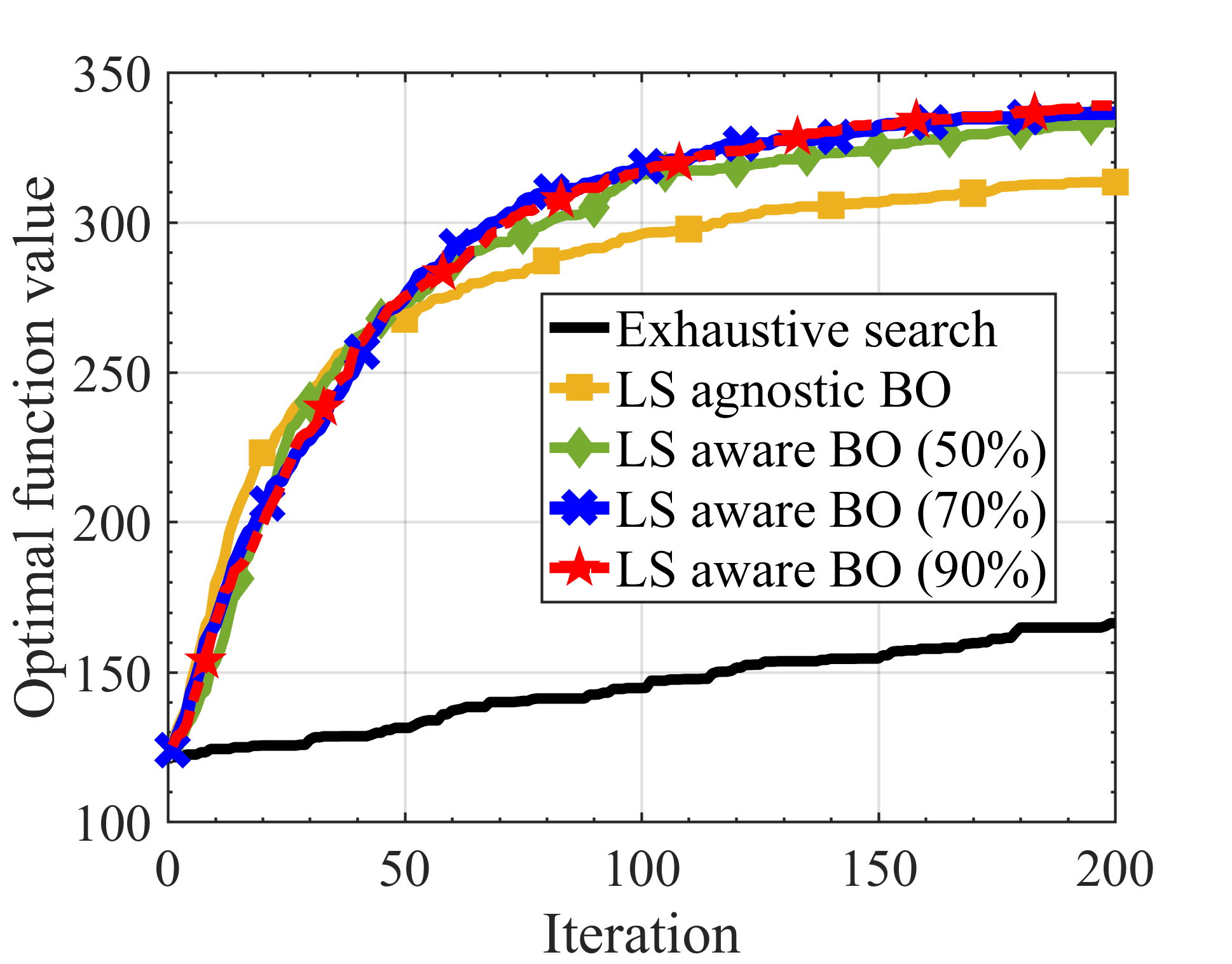}%
    \put(-5,72){\textbf{(a)}}%
    \end{overpic}%
    ~%
    \begin{overpic}[width=0.45\textwidth]{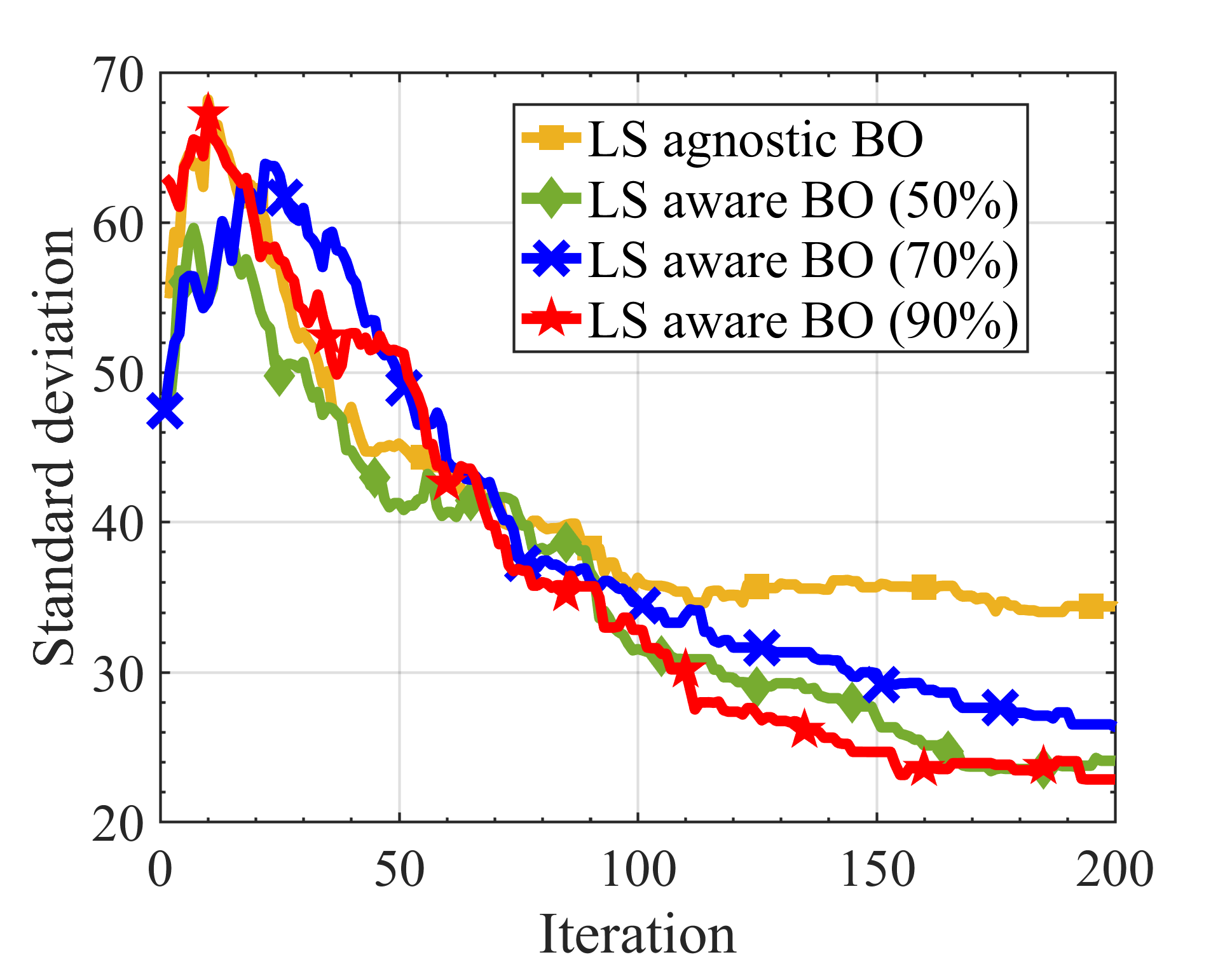}%
    \put(-5,72){\textbf{(b)}}%
    \end{overpic}%
    \\
    \begin{overpic}[width=0.45\textwidth]{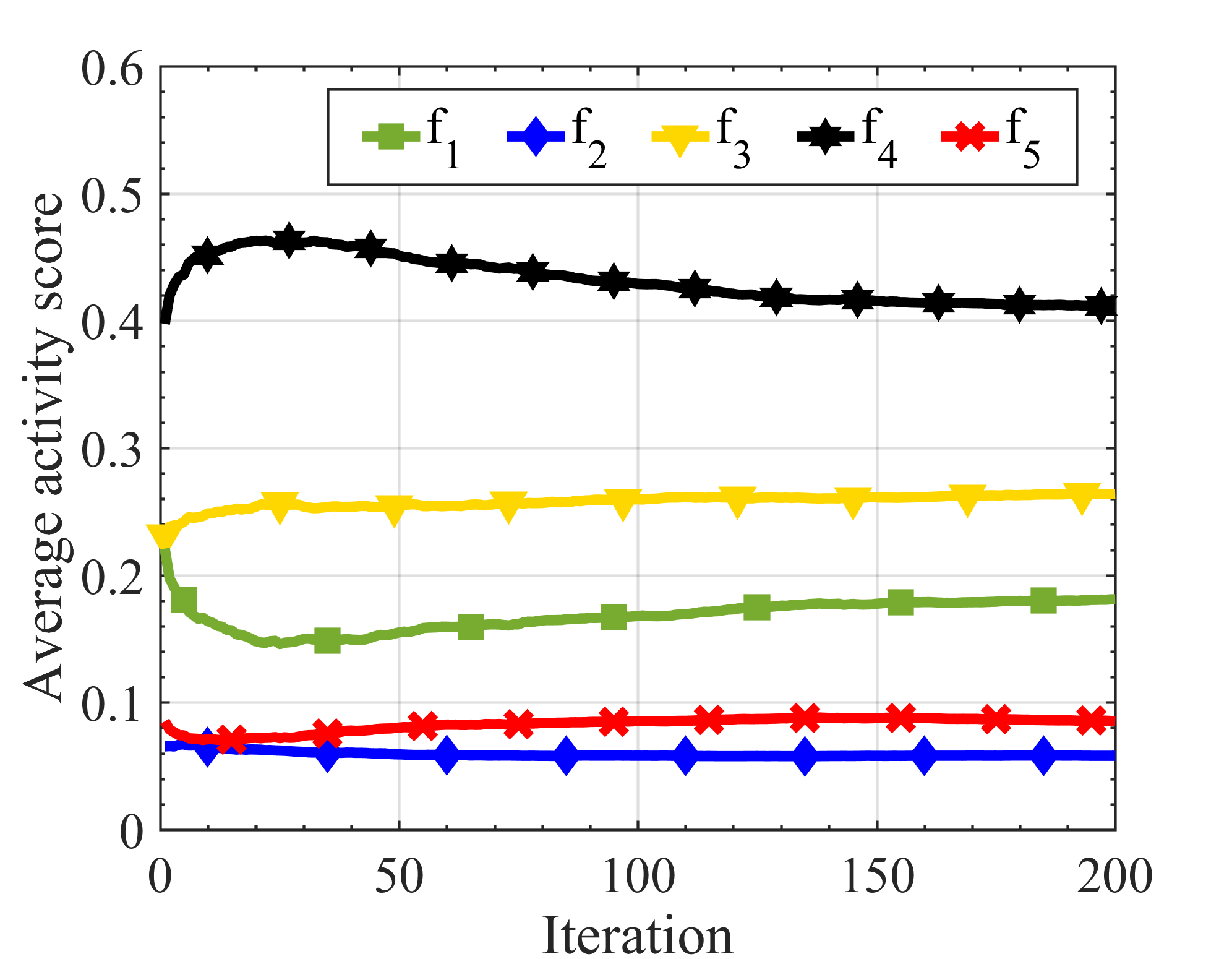}%
    \put(-5,72){\textbf{(c)}}%
    \end{overpic}%
    \begin{overpic}[width=0.45\textwidth]{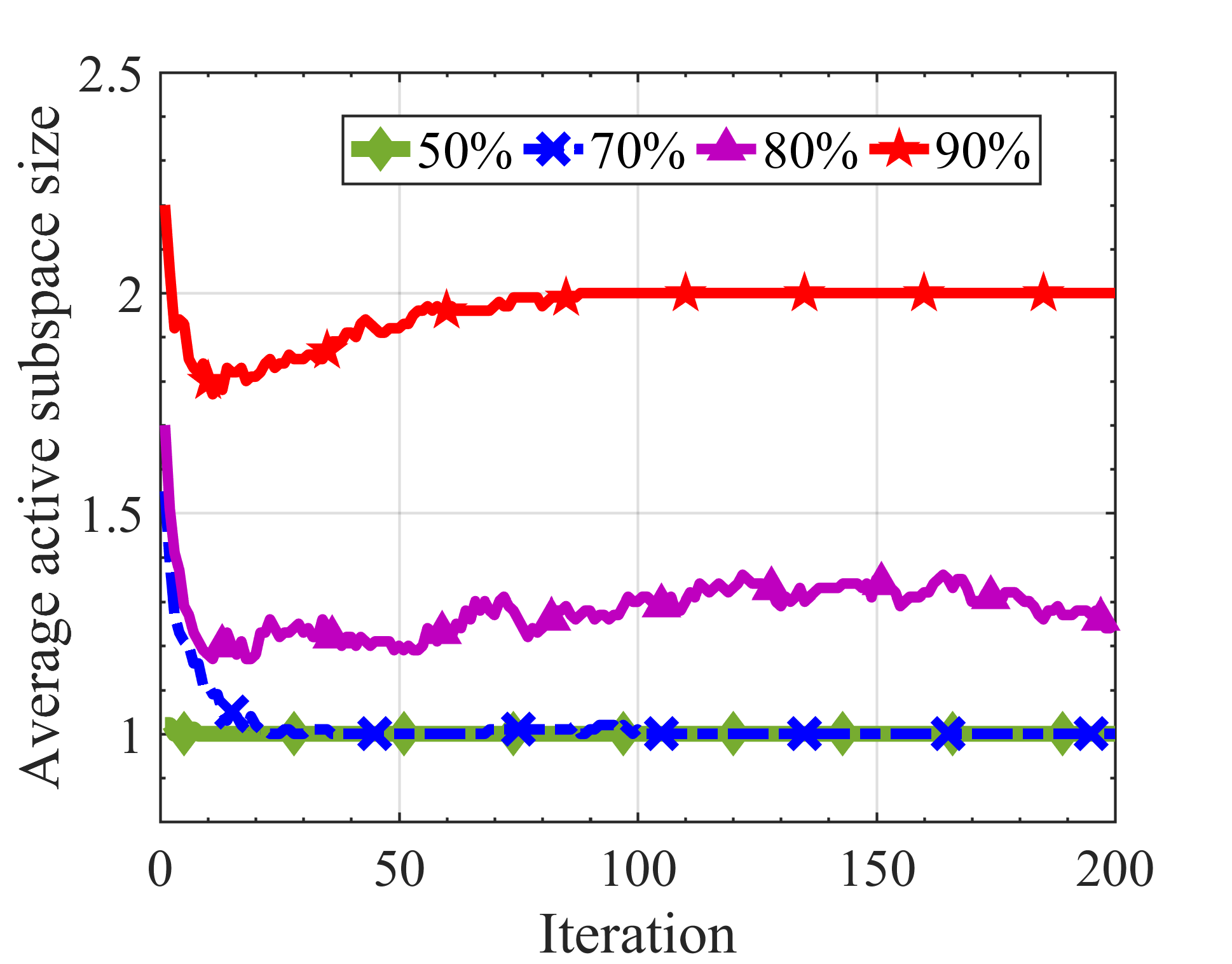}%
    \put(-5,72){\textbf{(d)}}%
    \end{overpic}%
    \caption{\textbf{Performance comparison of different optimization scenarios and the exhaustive search averaged over 100 replications of simulations.} The latent space-aware BO with different information-capturing thresholds outperforms the latent space-agnostic BO as higher function values are discovered with narrower confidence intervals. (a) The average discovered optimal function values and (b) the corresponding standard deviations. (c) \textbf{Activity scores as a metric of global sensitivity of the objective function to each latent variable.} The objective function is more sensitive to latent variables with larger activity scores. The plot shows that the framework is quickly recognizing the most important latent variables. (d) \textbf{Average active subspace size as functions of iterations.} A higher threshold requires constructing a higher-dimensional active subspace to capture more information about the relation between objective function variability and latent variables.}
    \label{toy_per}
\end{figure*}

In addition to studying the performance of the proposed design framework, we are also interested in investigating how much information from the latent space is optimal to capture. Thus, different information-capturing thresholds are tested: 50\%, 70\%, and 90\%. Capturing more information comes at the expense of increasing the size of the design space due to the formation of higher-dimensional active subspaces. To obtain the average performance of the design frameworks, simulations are repeated 100 times with different training datasets of 100 data points, generated by the Latin hypercube sampling technique. Figure~\ref{toy_per} shows (a) the average optimal function value and (b) the corresponding standard deviations as functions of iterations. The results are also compared against the exhaustive search scenario to emphasize the efficiency of the BO-based design frameworks. While both the latent space-agnostic and latent space-aware scenarios perform very similarly at the beginning stages, after a couple of iterations the latent space-aware BO starts to show its superiority. Such an improvement can be explained by noting that the active subspace is a gradient-based technique; thus, collecting more observations improves the accuracy of the active subspace calculations that determine the impact of different latent variables on the objective function variability. It is important to mention that all cases are expected to converge to the same optimal region; however, the latent space-agnostic BO needs to learn the complex relation between the input variables and the objective function solely based on input/output data.

An active subspace-based metric for global sensitivity analysis is the activity score formulated as:
\begin{equation}
    s_i = \sum_{j=1}^{n} \zeta_j w_{i,j}^2 , \quad i=1,...,m.
\end{equation}
Here, \(s_i\) is the activity score associated with the \(i\)th latent variable, and \(\zeta_j\) is the eigenvalue corresponding to the \(j\)th eigenvector \(w_j\). The more important a latent variable, the more it aligns with eigenvectors with larger eigenvalues, thus earning a higher activity score. Figure~\ref{toy_per}(c) shows the average activity scores of the 5 latent variables as functions of iterations. Note that the active subspace is dynamically updated as more observations are collected, and the activity scores vary accordingly during optimization. This plot demonstrates that not all latent variables have a similar impact on the objective function variability. In this synthetic example, the most important latent variable is \(f_4\), while \(f_2\) has the least impact on the objective function variability. Such sensitivity analysis is essential for informing designers about how latent variables can control a target property. A larger activity score indicates higher sensitivity of the objective function to a specific feature, and incorporating that feature into the design process can further accelerate convergence by providing critical information about the target property.

Figure~\ref{toy_per}(d) shows the average active subspace size for different information-capturing thresholds. As expected, on average, a higher threshold necessitates forming a higher-dimensional active subspace. It is observed that the first eigenvector is sufficient to capture at least 50\% of the variation in the objective function, as the size of the active subspace is consistently one. After several iterations and the accumulation of more observations, a single eigenvector can explain more than 70\% of the variation. However, if the goal is to capture at least 90\% of the variation, a second eigenvector should be included in the active subspace. The values in Fig.~\ref{toy_per}(d), averaged over 100 simulations, indicate that lower-dimensional scenarios occur more frequently. According to Fig.~\ref{toy_per}(c), latent variables \(f_2\) and \(f_5\) have the lowest activity scores, and their contributions in forming the active subspace are minimal unless the threshold is set very close to 1.

Using the synthetic function above, we simulate a condition where the goal is to optimize a material property while measurable microstructural features serve as the intermediate space connecting chemistry/process variables and the target property. We now apply the design framework to a more realistic design problem.

\subsection*{Demonstration on Chemistry–Microstructure–Property Modeling}

The framework is motivated by the prediction of the thermal conductivity (\(\kappa\)) of effective multiphase materials, which arises from the complex interactions of various energy carriers at the atomic level, including lattice vibrations, electrons, and photons.. While thermal analysis typically focuses on a dominant energy carrier, complexity arises in systems with multiple carriers, where \(\kappa\) comprises contributions from the lattice (\(\kappa_L\)), electrons (\(\kappa_e\)), and photons (\(\kappa_r\)), expressed as \(\kappa=\kappa_L+\kappa_e+\kappa_r\). This is particularly significant in thermoelectric materials, where both \(\kappa_L\) and \(\kappa_e\) play key roles~\cite{toberer2012advances}. Controlling thermal conductivity in multiphase materials, however, requires a holistic approach that considers intrinsic phase properties as well as microstructural and interfacial phenomena. Key strategies include engineering grain boundaries, interfaces, and defects; leveraging anisotropy and phase transformations; and tailoring chemical composition and managing thermal expansion mismatches~\cite{cahill1989thermal,lu2002size,vaqueiro2010recent,shakouri2011recent,bian2006beating}. Determining the maximum power-generation efficiency of isotropic thermoelectric materials involves factors like Carnot efficiency (\(\eta_\text{Carnot}=\frac{T^{hot}-T^{cold}}{T^{hot}}\)) and the generalized Zener criterion (\(Ze=(\sqrt{1+ZT}-1)/(\sqrt{1+ZT}+1)\))~\cite{he2017advances,rowe2018thermoelectrics}: 
    \begin{equation}
        \eta_{local}^{Max} = \eta_\text{Carnot} \, Ze \, \bigg[ \frac{\sqrt{1+ZT}-1}{\sqrt{1+ZT} + \Big(\frac{T^{cold}}{T^{hot}}\Big)} \bigg]
    \end{equation}
where \(Z\) is the thermoelectric figure of merit (\(Z=\frac{S^2\sigma}{\kappa}\)) derived using macroscopic heat balance in thermoelectric legs. In this context, Bian et al.~\cite{bian2006beating} demonstrated that heterogeneity is crucial in maximizing thermoelectric response, while Snyder et al.~\cite{snyder2003thermoelectric} proposed a compatibility factor for designing functionally graded thermoelectric materials. In this way, the complex problem of thermoelectric energy conversion has been reframed as optimizing a set of macroscopically measurable transport parameters (\(\sigma\), \(S\), and \(\kappa\)) simultaneously.

\begin{figure}
    \centering
    \includegraphics[width=0.90\columnwidth]{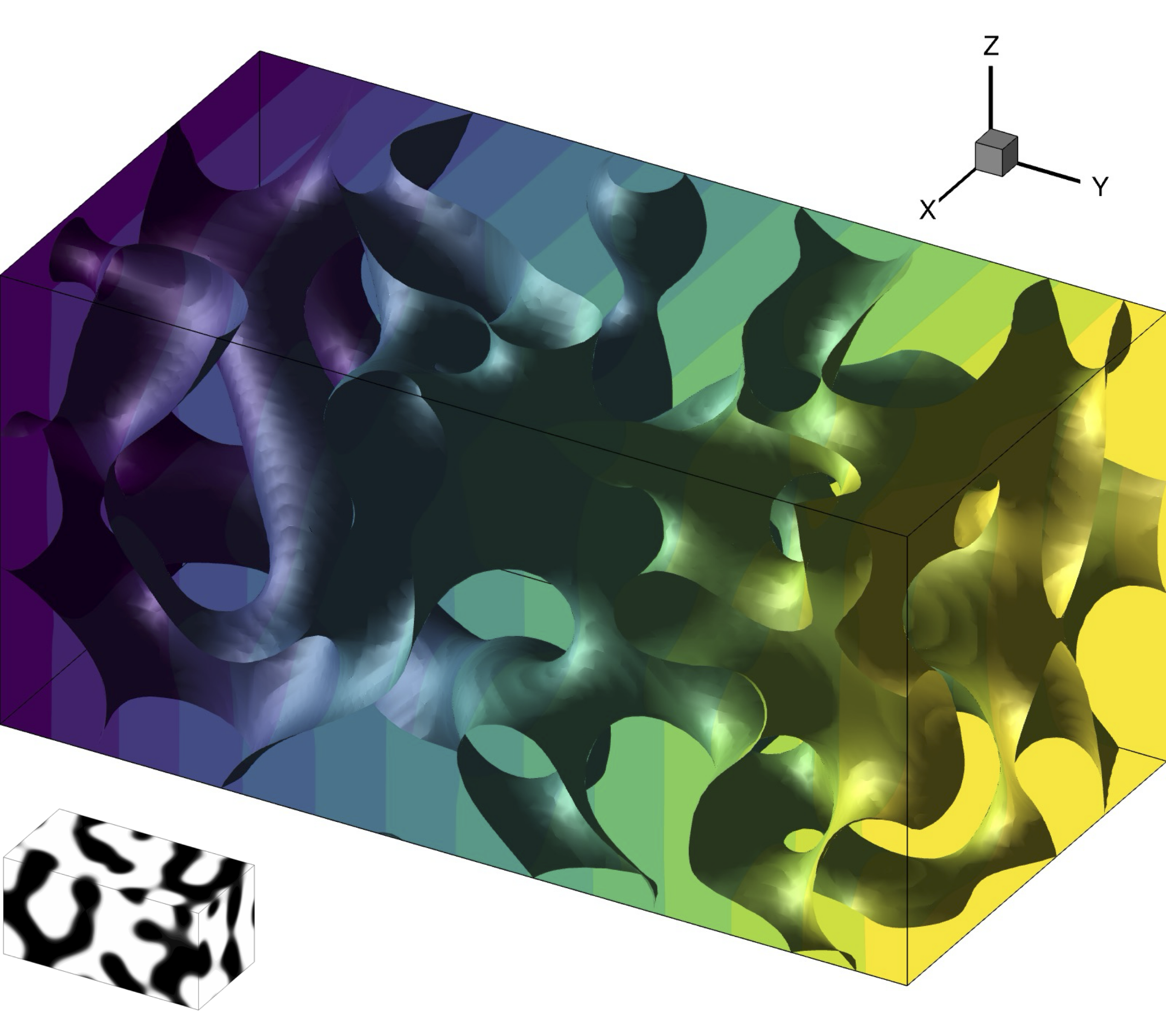}
    \caption{Thermal conduction across nanostructured microstructure grain boundaries, depicting heat gradients from one side to the other.}
    \label{fig:microstructure-}
\end{figure}

In pursuit of enhancing thermoelectric efficiency, minimizing thermal conductivity stands as a critical objective~\cite{rowe2018thermoelectrics}. By reducing heat transfer within a material, thermoelectric devices can better sustain the thermal gradient essential for effective energy conversion. Strategies such as nanostructuring, alloying, and phonon scattering are integral in reducing \(\kappa\), thereby optimizing the Seebeck coefficient and electrical conductivity in thermoelectric materials~\cite{snyder2003thermoelectric}. This concerted approach not only boosts device efficiency but also advances sustainable energy harvesting and waste heat recovery. A microstructure-sensitive thermal conductivity minimization problem is defined as:
\begin{equation}
\label{TC_minimzation}
   \text{Find} \quad \mathbf{x}^{*}= \arg\min_{\mathbf{x} \in \mathcal{X}}(\kappa)
\end{equation}
with a 4-dimensional design space, \(\mathbf{x}=[c_\text{Si}, \varepsilon^T_{ij}, C_{11}^{\text{Mg}_2\text{Si}}, C_{11}^{\text{Mg}_2\text{Sn}}]\) within \(\mathcal{X}\subset \mathbb{R}\), mapped to a 5-dimensional microstructure latent space, \(\mathbf{F}=[A_f, c_{\text{Mg}_2\text{Sn}}, c_{\text{Mg}_2\text{Si}}, PS^{max}, S^{shannon}]\), and then minimizing the objective property \(\kappa\) as a function of latent variables. It is worth noting that Mg\(_2\)Sn\(_{1-x}\)Si\(_x\) forms a cubic structure, which allows the simplification of thermal conductivity to a scalar value.

In determining material properties, higher-order microstructural attributes such as continuity exert significant influence. A detailed theoretical understanding of random heterogeneous materials relies on precisely characterizing microstructural details including phase volume fractions, interface surface areas, orientations, sizes, shapes, spatial distributions, connectivity, and related factors. It has been established that the thermal conductivity of a heterogeneous periodic medium behaves similarly to that of a homogeneous medium—governed by the steady-state conduction equation with a constant conductivity tensor in the limit of rapid microstructural fluctuations~\cite{torquato2002random}. Initially, we compile a comprehensive dataset of microstructures through high-throughput phase-field simulations. These simulations generate time-series data depicting the microstructures of Mg\(_2\)Sn\(_{1-x}\)Si\(_x\) alloy for various compositions, enabling the study of uncertainty propagation in the microstructure during isothermal thermal annealing. Subsequently, we assess the effective thermal conductivity of these microstructure evolutions.

\begin{figure*}[!ht]
    \centering
    \begin{overpic}[width=0.45\textwidth]{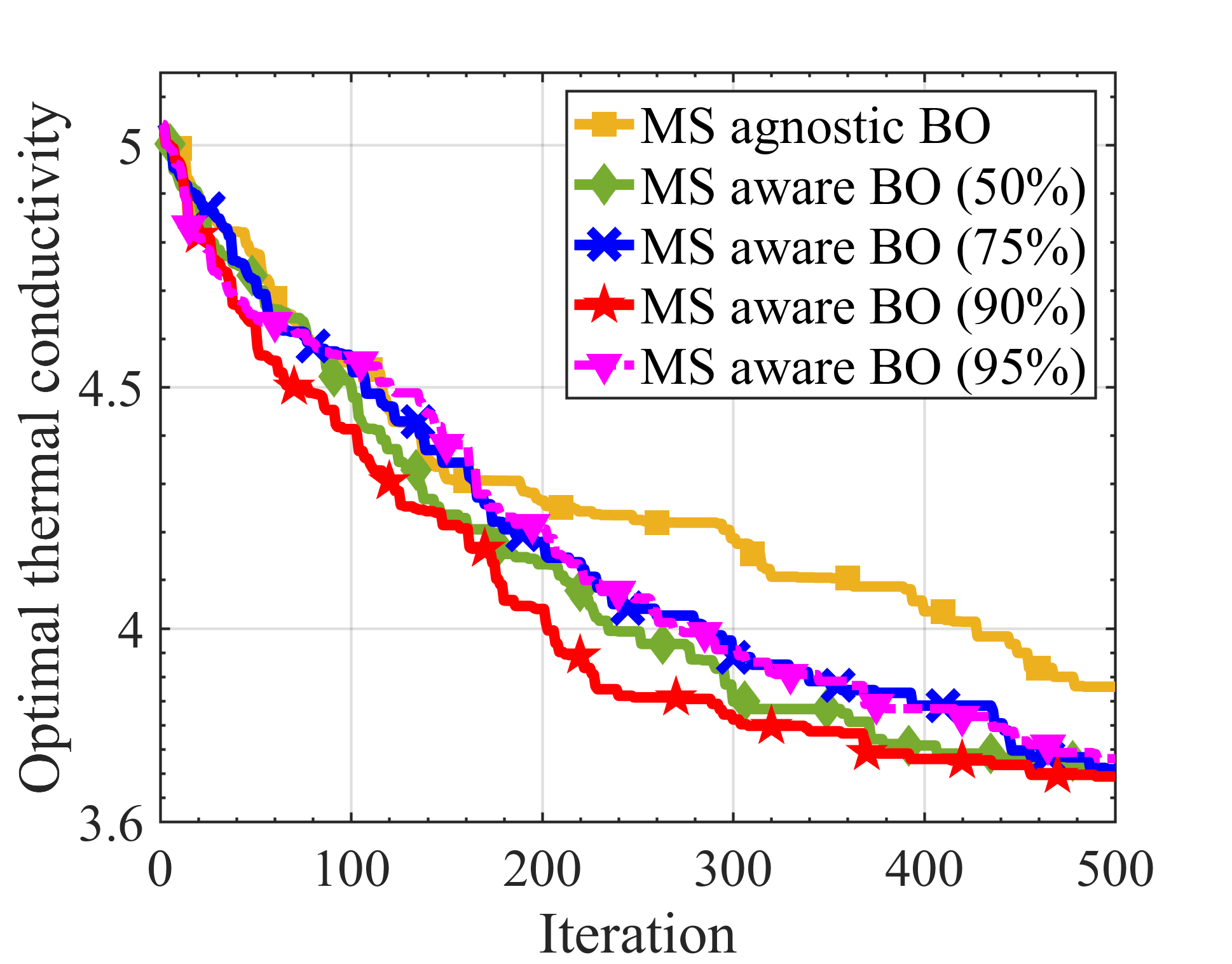}%
    \put(-5,72){\textbf{(a)}}%
    \end{overpic}%
    ~%
    \begin{overpic}[width=0.45\textwidth]{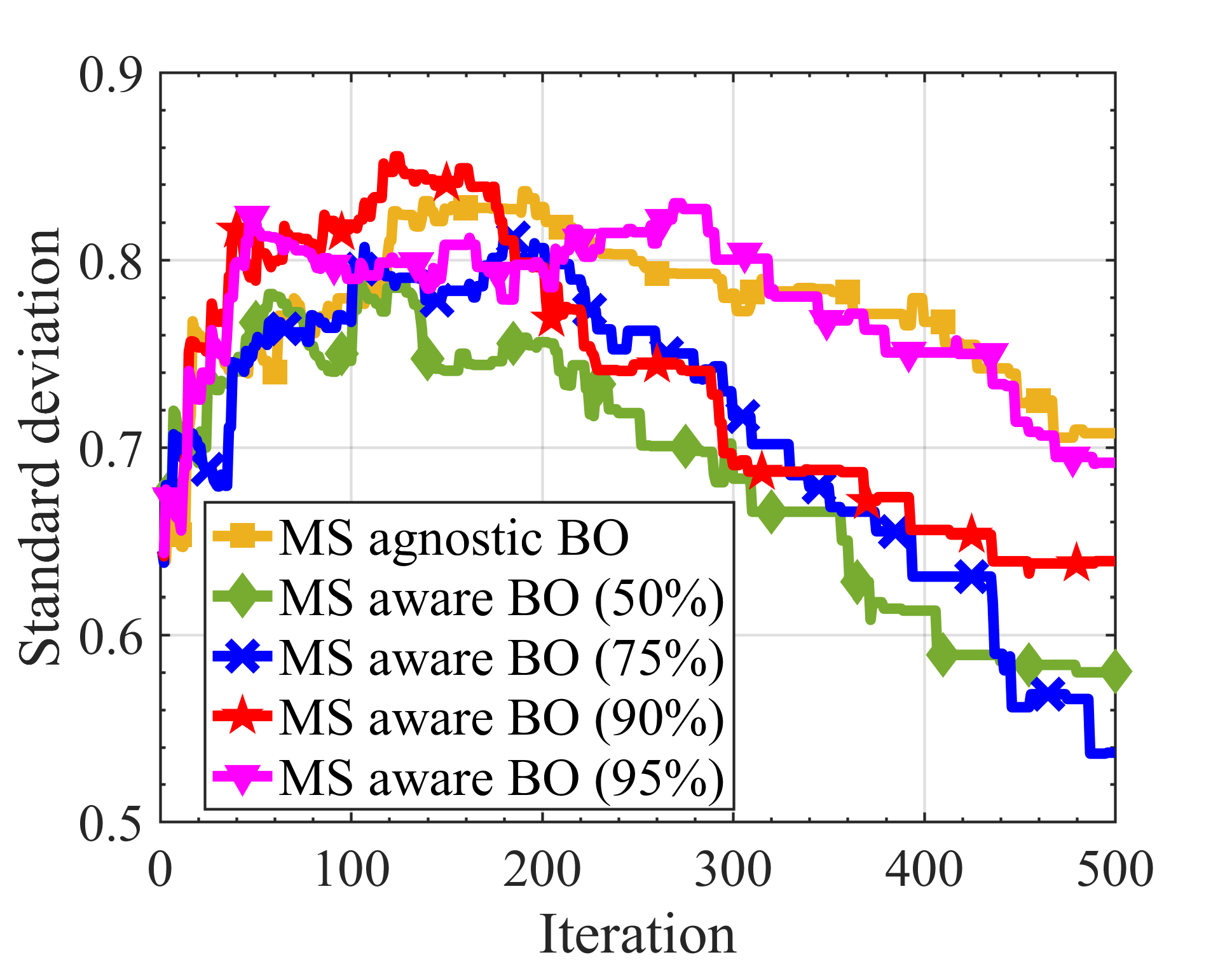}%
    \put(-5,72){\textbf{(b)}}%
    \end{overpic}%
    \\
    \begin{overpic}[width=0.45\textwidth]{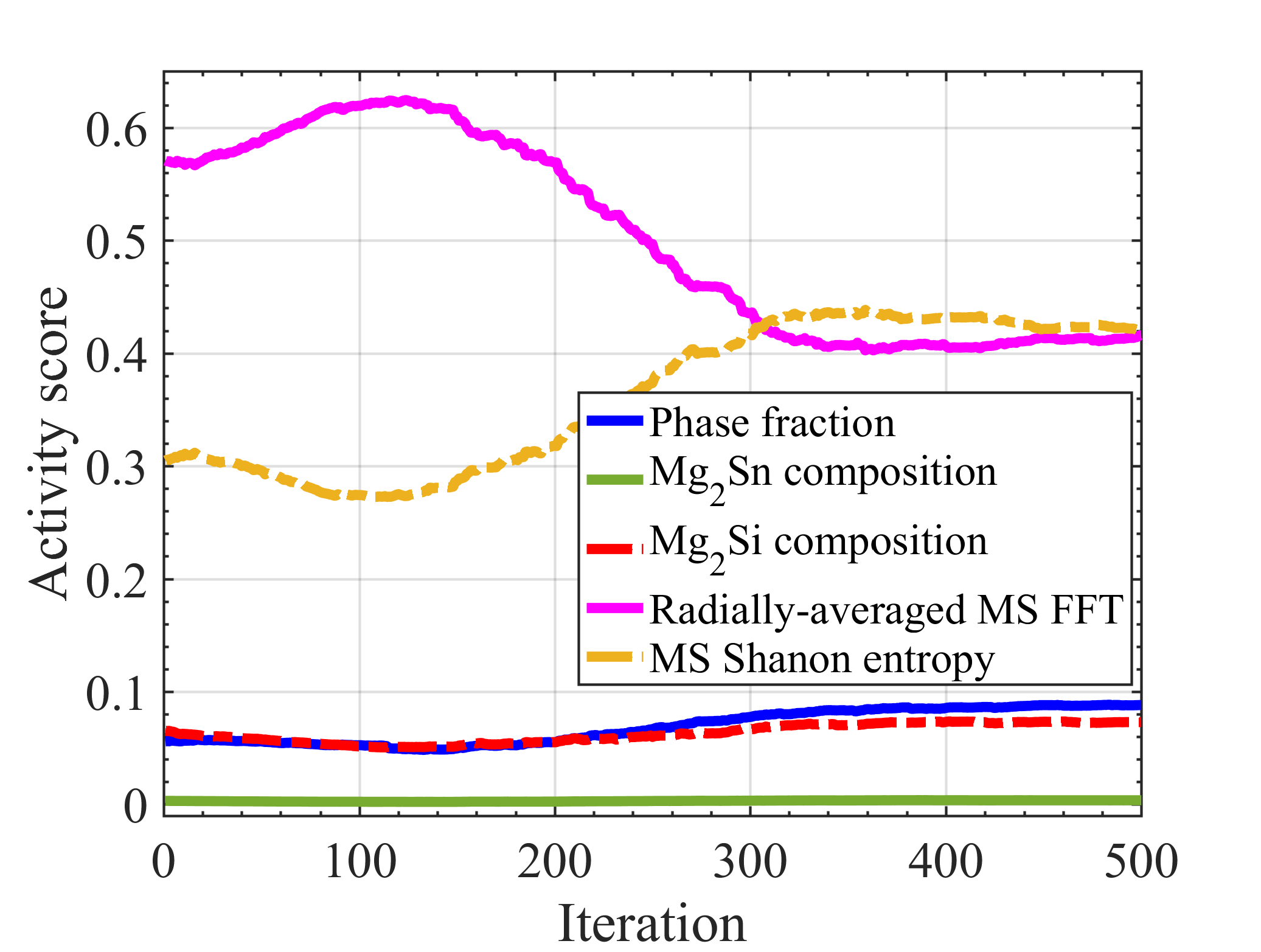}%
    \put(-5,72){\textbf{(c)}}%
    \end{overpic}%
    \begin{overpic}[width=0.45\textwidth]{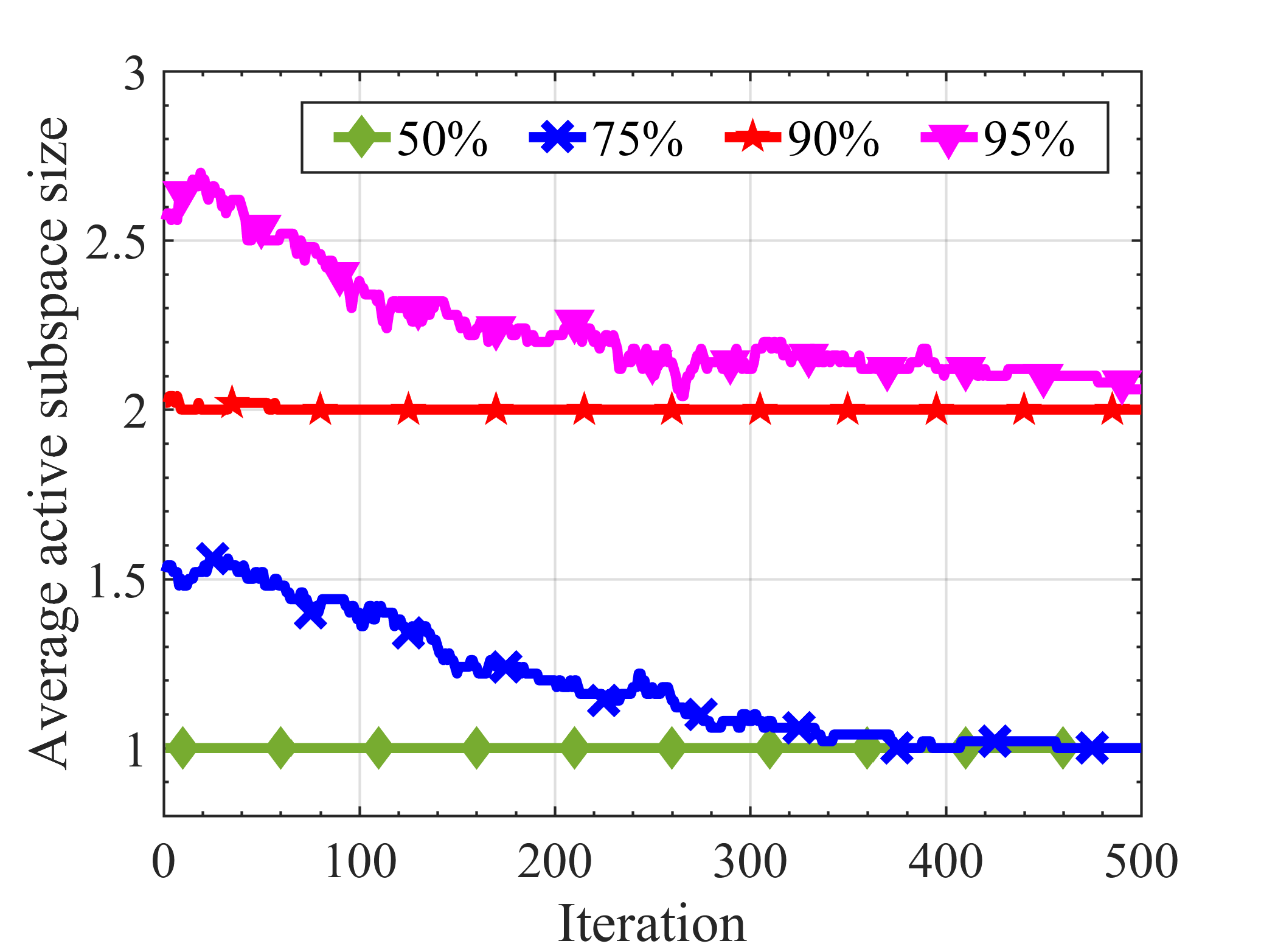}%
    \put(-5,72){\textbf{(d)}}%
    \end{overpic}%
    \caption{(a) and (b) \textbf{Optimum thermal conductivity obtained using different optimization scenarios.} Microstructure-aware optimization scenarios are run with different variation-capturing thresholds of 50\%, 75\%, 90\%, and 95\%. (a) The average thermal conductivity values and (b) the corresponding standard deviations. (c) \textbf{The average activity scores as measures of sensitivity to microstructure features.} The analysis indicates that thermal conductivity is more sensitive to changes in the radially averaged microstructure FFT and image entropy. The sensitivity to these features is similar around the optimum design region. (d) \textbf{The average active subspace dimensionality as functions of iterations.} To capture more variation of the objective function in the feature space, a higher-dimensional active subspace is required. Toward the end of the design process, lower-dimensional active subspaces are sufficient to explain variations in thermal conductivity with changes in microstructure descriptors.}
    \label{per}
\end{figure*}

We assess the framework using the chemistry–microstructure–property dataset derived from high-throughput microstructure modeling of Mg\(_2\)Sn\(_{1-x}\)Si\(_x\) alloy. Figure~\ref{per}(a) and (b) display the average thermal conductivity values and the corresponding standard deviations during minimization, evaluated under various information-capturing thresholds (50\%, 75\%, 90\%, and 95\%). The results demonstrate that while both latent space-agnostic and latent space-aware BO exhibit similar performance initially, the latent space-aware BO outperforms the agnostic approach after approximately 200 iterations. All cases ultimately converge to a similar region in the thermal conductivity domain. The latent space-agnostic BO must learn the intricate relationship between input variables and the objective function solely from direct chemistry/property data, whereas the latent space-aware BO leverages microstructural features as latent variables to achieve optimal designs more efficiently.

Figure~\ref{per}(c) illustrates the average activity scores of the five latent variables (i.e., phase fraction, Mg\(_2\)Sn phase composition, Mg\(_2\)Si phase composition, radially averaged FFT structure function, microstructure Shannon entropy) as a function of BO iterations. Notably, as the active subspace is dynamically updated with more observations, the activity scores vary throughout the optimization. This analysis demonstrates that not all latent variables contribute equally to the objective function variability; in this example, the radially averaged FFT structure function and Shannon entropy are the most influential, jointly explaining nearly 90\% of the total activity score.

Figure~\ref{per}(d) shows the average size of the active subspace for different information-capturing thresholds in the context of minimizing thermal conductivity. As anticipated, a higher threshold necessitates forming a higher-dimensional active subspace. For instance, capturing 50\% of the variation is achieved with a one-dimensional active subspace, while a second eigenvector is required to capture at least 90\% of the variation. These observations, averaged over 100 simulations, indicate that lower-dimensional subspaces occur more frequently under lower thresholds. According to Fig.~\ref{per}(c), latent variables \(f_2\) and \(f_5\) have the lowest activity scores, contributing minimally to the active subspace unless the threshold is set very close to 1.

In both the synthetic and material design examples, the results indicate that incorporating latent space information improves design performance by providing valuable insight into the variability of the objective function. The latent space-agnostic BO relies solely on direct input–output relationships, whereas the latent space-aware BO exploits intermediate microstructural information, which, being dependent on the design variables, enriches the mapping between design space and the objective function—often requiring fewer observations. Furthermore, the use of the active subspace technique facilitates on-the-fly sensitivity analysis, informing designers about the most critical structural features that, when controlled, can further accelerate performance improvements.

In some cases, the property may be measured directly without the need for intermediate structural information. For example, if the goal is to optimize the yield strength of an alloy, a tension test might suffice. In such scenarios, measuring microstructure characteristics could impose additional costs. A future study should investigate the trade-off between the extra cost of measuring structural features and the resource savings achieved by reducing the number of experiments. Moreover, multi-objective optimization—a common design scenario where multiple conflicting performance metrics must be balanced—might require the simultaneous exploitation of information from multiple latent spaces. This extension represents another promising direction for future investigation.

\section{Conclusions}
\label{sec:conclusions}

In this work, we introduced a microstructure (or latent space)-aware Bayesian optimization (BO) framework that explicitly treats microstructural features as design parameters. By integrating Gaussian process (GP) regression with the active subspace method, our approach captures the high-dimensional process–structure–property relationship while quantifying uncertainty in data-limited settings. This integrated framework reveals how microstructural descriptors can be tuned to achieve target materials properties more efficiently than conventional methods that treat microstructure only as an emergent outcome.

Our results, demonstrated on both synthetic examples and a real-world computational case study, show that reducing the dimensionality of complex latent spaces via ASM improves the accuracy of the GP surrogate. This, in turn, facilitates a more precise mapping from processing parameters to material properties. Importantly, these findings suggest that a shift from traditional approaches—where microstructure is observed only after fabrication—to one where microstructure is an explicit addressable variable can offer more effective and efficient pathways to accelerated materials development.

The proposed framework also holds promise for future autonomous materials discovery. In Self-Driving Laboratories (SDLs), real-time or offline microstructural characterization could provide continuous feedback, further refining the surrogate model used for the optimal inversion of PSPP linkages. Such integration would enable adaptive, closed-loop optimization that navigates high-dimensional microstructure-aware design spaces more rapidly and with higher precision.

Future work could extend this framework to handle multi-objective optimization and incorporate more advanced physics-based models. Although the current approach leverages sparse data through a fully probabilistic treatment of the process–structure–property chain, challenges remain in scaling the method for even more complex systems. Addressing these challenges will further enhance the ability to deploy microstructure-sensitive alloy design under increasingly realistic settings.

\section*{Declaration of Competing Interest}
The authors declare that they have no known competing financial interests or personal relationships that could have appeared to influence the work reported in this paper.

\section*{Acknowledgements}
Research was sponsored by the Army Research Laboratory and was accomplished under Cooperative Agreement Number W911NF-22-2-0106. The views and conclusions contained in this document are those of the authors and should not be interpreted as representing the official policies, either expressed or implied, of the Army Research Laboratory or the U.S. Government. The U.S. Government is authorized to reproduce and distribute reprints for Government Purposes, notwithstanding any copyright notation herein. The authors also acknowledge the support of the National Science Foundation (NSF) through Grant No. CDSE-2001333 and the Army Research Office (ARO) through Grant No. W911NF-22-2-0117. The authors also acknowledge the computational resources provided by Texas A\&M High-Performance Research Computing (HPRC) Facility, as well as Texas A\&M's Vice-President for Research for their support through the ASCEND-TPT program.

\section*{Authors' Contributions}
\noindent
\textbf{D. Khatamsaz:} Conceptualization, implementation, writing (original draft), editing. \\
\textbf{V. Attari:} Conceptualization, implementation, writing (original draft), editing. \\
\textbf{R. Arr\'{o}yave:} Conceptualization, funding acquisition, writing (review and editing).

\section*{Data Availability}
Data generated from the Bayesian optimization framework are available from the corresponding author upon reasonable request.

\bibliographystyle{elsarticle-num}

\bibliography{Refs_Danial,Refs_Vahid, Refs_Ray}

\end{document}